# Investment behavior and firms' financial performance: A comparative analysis using firm-level data from the wine industry

Claudiu Tiberiu ALBULESCU[a,b*]

[a] *Management Department, Politehnica University of Timisoara, 2, P-ta. Victoriei, 300006 Timisoara, Romania*
[b] *CRIEF, University of Poitiers, 2 rue Jean Carbonnier, 86022 Poitiers, France*

**Abstract**
This paper assesses the role of financial performance in explaining firms' investment dynamics in the wine industry from the three European Union (EU) largest producers. The wine sector deserves special attention to investigate firms' investment behavior given the high competition imposed by the latecomers. More precisely, we investigate how the capitalization, liquidity and profitability influence the investment dynamics using firm-level data from the wine industry from France (331 firms), Italy (335) firms and Spain (442) firms. We use data from 2007 to 2014, drawing a comparison between these countries, and relying on difference- and system-GMM estimators. Specifically, the impact of profitability is positive and significant, while the capitalization has a significant and negative impact on the investment dynamics only in France and Spain. The influence of the liquidity ratio is negative and significant only in the case of Spain. Therefore, we notice different investment strategies for wine companies located in the largest producer countries. It appears that these findings are in general robust to different specifications of liquidity and profitability ratios, and to the different estimators we use.

---

[*] Correspondence: Claudiu Tiberiu Albulescu, Politehnica University of Timisoara, P-ta. Victoriei, No. 2, 300006, Timisoara, Romania. Tel: 0040-743-089759. Fax: 0040-256-403021. E-mail: claudiu.albulescu@upt.ro, claudiual@yahoo.com. http://orcid.org/0000-0003-2875-1749.

## 1. Introduction

One of the key challenges the wine economics and corporate finance literature has to cope with is the identification of determinants of firms' investment behavior. Understanding the factors influencing firms' investment is important both from the point of view of business cycle fluctuations, and from the perspective of financial management optimization and investors' wealth. For this purpose, prior literature investigates the role of a large set of external and internal determinants, and reports mixed empirical evidence. However, the interest for studding the investment behavior of wine companies is scarce. This paper fills in this gap and adds to the menu of studies addressing the role of internal factors in supporting the firms' investment behavior, by focusing on the role of financial performance and using wine industry firm-level data from the largest wine producing countries, namely France, Italy and Spain. We posit that the investment behavior of the wine companies located in these countries is not only influenced by the economic context and competition policies (Rizzo, 2019), but also by their financial performances.

The external determinants of firms' investment behavior are related to business cycle (Gertler and Gilchrist, 1994; Jeon and Nishihara, 2014; Pérez-Orive, 2016), taxation (Hall and Jorgenson 1967; Morck, 2003; Jugurnath et al., 2008), monetary policy (Vithessonthi et al., 2017), quality of institutions (Ajide, 2017), and even to the behavior of other firms from the same industry (Lyandres, 2006; Leary and Roberts, 2014; Park et al., 2017). Noteworthy studies (e.g. Abel 1983; Bernanke 1983; Hartman 1972; Pindyck 1988; Calcagnini and Iacobucci 1997; Baum et al., 2008; Glover and Levine, 2015) investigate the controversial role of uncertainty in influencing firms' investment behavior.[1]

Two main categories of internal factors explain firms' investment behavior.[2] On the one hand, building upon Modigliani and Miller (1958), the literature underlines the role of financial constraints, leverage and cash flow (Fazzari et al., 1988; Gilchrist and Himmelberg, 1995; Lang et al., 1996; Chen et al., 2001; Suto, 2003; Aivazian et al., 2005;  Ahn et al., 2006; Baum et al., 2010; Almeida et al., 2011; Maçãs Nunes et al., 2012; Colombo et al., 2013; Vermoesen et al., 2013; Ameer, 2014). On the other hand, agency costs, information asymmetry and ownership structure are put forward (Jensen and Meckling, 1976; Koo and Maeng, 2006; Danielson and

---

[1] Uncertainty is in general associated with the lack of forecast accuracy (Albulescu et al., 2017). A recent paper by Chen et al. (2017) shows that the quality of analysts' forecasts significantly increases the efficiency of firms' investment.

[2] A distinct category of internal factors explaining firms 'investment behavior might be related to the technological capabilities (for a discussion, please see the recent paper by Kang et al., 2017).

Scott, 2007; Alex, et al., 2013; Farla, 2014; Mavruk and Carlsson, 2015). Several papers (e.g. Shen and Wang, 2005) show that both financial constraints and ownership structure influence the investment decision, while other papers (e.g. Bokpin and Onumah, 2009) underline the role of firms' size in explaining the investment behavior.

The financial constraints and firms' leverage have important implications on the investment behavior (Suto, 2003; Ahn et al., 2006), influencing at the same time the structure of investment (Almeida et al., 2011). A series of studies shows that financial constraints have a negative impact on firm-level investment. In this line, Vermoesen et al. (2013) report that high leveraged Belgian firms experienced a larger investment contraction during crisis times, as compared to less leveraged firms. Opposite findings are reported by Baum et al. (2010) for a set of manufacturing United States (US) firms, who show that leverage stimulates the investment under the effects of uncertainty. However, most of existing empirical works focus on the role of financial constraints in explaining the investment – cash flow sensitivities. The financial friction theory mentions that the impact of cash flow on investment increases in the presence of credit constraints. While Aidogan (2003) shows that the sensitivity of firm's investment to its own cash flow increases for growing firms, Kim (2014) states that the investment – cash flow sensitivity is explained by the level of external financing. Using a Panel Smooth Transition Regression model for 519 Asian listed firms over the period 1991-2004, Ameer (2014) reports that investment – cash flow sensitivity varies across different categories of firms. Mulier et al. (2016) also point out that the highest investment – cash flow sensitivity characterizes financially constrained firms. Another set of works (e.g. Gamba and Triantis 2008; Arslan-Ayaydin et al., 2014) underlines the role of financial flexibility in fostering firm-level investment. Using a sample of 1,068 Asian firms, Arslan-Ayaydin et al. (2014) report that financial flexibility achieved through conservative leverage policies has significant influence on investment, in particular in crisis periods.

The second strand of literature investigates the role of agency costs, information asymmetry and ownership structure in influencing the investment behavior. In their pioneering paper, Jensen and Meckling (1976) show that agency conflicts might distort firms' investment decision in the presence of multiple owners. Performing an empirical investigation for a panel of 115 listed firms in Taiwan for the period 1991-1997, Shen and Wang (2005) highlight that investment behavior is financially constrained in a cross-ownership system. At the same time, Koo and Maeng (2006) find that the presence of foreign ownership in Korean firms decreases the investment – cash flow sensitivity. More recently, Farla (2014) discovers that firms' investment behavior has little dependency on a country's macroeconomic setting, while foreign-owned firms have lower investment dynamics.

Only few papers, however, focus on the role of profitability and liquidity on the investment behavior (e.g. Perić and Đurkin, 2015; Yu et al., 2017). While some studies (e.g. Stickney and McGee, 1982; Gilchrist and Himmelberg, 1995; Black et al. 2000) use financial performance indicators as control variables in their empirical specifications, several papers put accent on the role of liquidity in influencing the investment behavior. As Baum et al. (2008) show, the impact of liquidity on investment is not straightforward. While in crisis periods characterized by credit contractions and financial frictions it is expected that liquidity positively influence the investment decision, an opposite effect appears if investment projects are delayed. On the one side, Acharya et al. (2007) state that the liquidity level sustains firms' future investment and offers protection against market risks. On the other side, Hirth and Viswanatha (2011) find that in the case of financially constrained firms, the relationship between liquidity and investment is U-shaped.

We extend the existing literature by examining not only the role of liquidity, but also the impact of capitalization and profitability on investment behavior. All these variables characterize the firms' financial performance, offering at the same time information about risk protection and incentive to develop the business. The level of cash holdings and thus the level of liquidity is considered the cheapest cost of investment. Therefore, for a specific period, if firms decide to increase their liquidity for risk protection reasons (i.e. during crisis periods), a trade-off is expected between liquidity and investment. The increase of capitalization level might also be done in the detriment of investment. It is surprising that previous literature does not debate the role of capitalization in the investment behavior. However, the level of capitalization provides, on the one hand, information about the debt level and, on the other hand, information about the way shareholders interact with managers in the investment decision. When investment becomes risky, shareholders might prefer to increase capitalization. At the same time, shareholders' equity represents an investment resource. In this context, during a fiscal year, it is expected that an increase in capitalization negatively influence the investment dynamics. Finally, the level of profitability positively affects the investment behavior. First, profitability increases the level of internal funds available for investment and has a negative influence on leverage (Datta and Agarwal, 2014). Second, high profits provide information about market dynamics and recommend future investments.

Another contribution of this paper to the bulk of literature investigating the determinants of firm-level investment consist in the empirical approach we use. Investment dynamics affects in its turn firms' financial performance (Gatchev et al., 2009). Therefore, in line with other studies, we address the endogeneity issues resorting to a Generalized Method of Moments (GMM) panel approach. Nevertheless, different form previous works, we address different econometric issues as residual autocorrelation or instruments' over-identification, which may introduce a bias in the reported results, if the models are not correctly specified. Comparing a difference-GMM (Arellano and Bond, 1991) and a system-GMM estimator (Blundell and Bond, 1998), we show that the results are sensitive to different econometric specifications, although they are robust to alternative measures of liquidity and profitability.

Finally, we investigate the role of financial performance on the investment behavior using wine industry firm-level data from France, Italy and Spain, the largest European Union (EU) and worldwide producers. As far as we know, the study by Outreville and Hanni (2013) is the only one addressing the determinants of investment in the wine industry. However, the authors focus on the foreign investment, investigating the case of the largest multinational enterprises, and underline the role of location for the inward investment. Different from this work, we analyze the case of domestic and foreign firms acting in the wine industry from the largest producing countries. France and Italy dominated the international wine market before the 1980s (Morrison and Rabellotti, 2017). Spain recorded a considerable development of the wine industry since then. Therefore, even after the increasing importance of newcomers in the industry (i.e. US, Chile, South Africa or Australia), the three EU countries continued to dominate the wine industry at global level.[3] Has the financial performance of firms located in these countries a similar impact on their investment behavior in the context of an increased competition on the wine market? We try to respond to this question analyzing firm-level data for 331 firms located in France, 335 firms located in Italy and 442 firms from Spain, over the period 2007 to 2014.

The rest of the paper is structured as follows. Section 2 presents some general statistics about the wine industry, with a focus on the EU. Section 3 describes the data and the methodology. Section 4 highlights the empirical results and presents the robustness checks. In Section 5 we present the summary of results and discuss in a comparative manner the role of financial performance on firms' investment behavior in the three analyzed countries, generating policy recommendations. The last section concludes.

---

[3] The EU countries does not only represent the largest wine exporters. For example, the United Kingdom counts between the largest wine importers (Anderson and Wittwer, 2017).

## 2. General statistics about the wine industry in the selected EU countries

During the last decades, in the context of new EU regulations, wine-producing regions of Europe struggled to adapt to changing market conditions and to fight against the competition of newcomers in this industry (Outreville and Hanni, 2013). Table 1 indicates that France, Italy and Spain together represented more than 55% from the total wine production, and more than 25% of total wine exports during the 1960s. However, the total production of these countries dropped to 45% out of the world production during the 2010s, while the total exports represent nowadays more than 50%. These figures show that world-level production and consumption increased with the newcomers on the wine market, but the consecrated producers became more and more competitive. This happened in the context of an intensive process of international acquisitions, driven by competitive prices and the opportunity to acquire key brands (Anderson et al., 2003). Given that wine is considered a typical cultural commodity, these producers readapted their market strategy, underlining the intangible characteristics of their product (e.g. the notion of 'terroir' in France). Nevertheless, while Italy and Spain continued to increase their quotas in the world exports, France encountered a severe contraction during the last decade.

Table 1. Wine production and exports (% world total volumes)

|     | 1961 | 1970 | 1980 | 1990 | 2000 | 2007 | 2008 | 2009 | 2010 | 2011 | 2012 | 2013 |
|-----|------|------|------|------|------|------|------|------|------|------|------|------|
| Wine production | | | | | | | | | | | | |
| FR  | 22.59% | 24.97% | 19.79% | 22.98% | 20.32% | 17.80% | 15.69% | 17.47% | 16.77% | 18.69% | 16.17% | 14.67% |
| IT  | 24.42% | 22.81% | 24.57% | 19.24% | 19.10% | 15.47% | 16.15% | 16.22% | 16.54% | 14.87% | 14.70% | 15.39% |
| SP  | 9.39% | 8.48% | 12.03% | 13.92% | 14.54% | 13.30% | 13.73% | 12.14% | 13.36% | 12.33% | 11.95% | 15.75% |
| Wine exports | | | | | | | | | | | | |
| FR  | 14.72% | 11.26% | 19.58% | 28.19% | 22.07% | 16.34% | 15.17% | 13.66% | 14.12% | 14.30% | 14.87% | 14.52% |
| IT  | 6.87% | 15.25% | 33.49% | 29.55% | 23.20% | 21.12% | 20.91% | 22.79% | 23.26% | 23.70% | 21.08% | 20.31% |
| SP  | 5.48% | 9.03% | 12.22% | 10.80% | 12.01% | 16.32% | 17.66% | 16.98% | 18.37% | 21.81% | 20.31% | 17.96% |

*Note: France (FR), Italy (IT), Spain (SP).*
Source: Faostat database

As compared to other EU countries, France, Italy and Spain are considered by far the largest producers, representing according to the Eurostat statistics, more than 80% of the total wine production in the EU. Table 2 presents the dynamics of the wine industry in terms of opening stocks in the selected EU countries.

Table 2. Opening stocks by vintage year in the EU countries (1,000 Hl)

|       | 2007-08 | 2008-09 | 2009-10 | 2010-11 | 2011-12 | 2012-13 | 2013-14 | 2014-15 | 2015-16 | 2016-17 |
|-------|---------|---------|---------|---------|---------|---------|---------|---------|---------|---------|
| FR    | 57,062  | 57,459  | 53,901  | 54,061  | 54,518  | 59,958  | 53,238  | 47,830  | 50,318  | 51,514  |
| IT    | 41,120  | 41,719  | 44,746  | 41,360  | 41,502  | 40,632  | 36,500  | 45,250  | 41,276  | 42,692  |
| SP    | 33,817  | 34,168  | 36,962  | 36,446  | 34,169  | 28,677  | 29,311  | 36,619  | 33,730  | 30,701  |
| EU-28 | 165,624 | 167,871 | 174,182 | 170,454 | 164,921 | 160,483 | 150,868 | 164,249 | 162,908 | 163,586 |

*Note: France (FR), Italy (IT), Spain (SP), European Union with 28 member states (EU-28).*
Source: Eurostat database

## 3. Data and methodology

*3.1. Data*

We use firm-level annual data from AMADEUS database to investigate the impact of firms' financial performance on the investment dynamics over the period 2007 to 2014. To avoid the broken panel bias, we have included in our analysis only firms without missing values for a specific indicator. Further, we have dropped from our sample those companies where data indicate a capitalization ratio (capital to total assets) over 100%. Finally, our sample includes 331 firms out of 367 firms registered in France (90%), 335 firms out of 410 recorded in Italy (82%), and 442 firms out of 531 registered in Spain (83%). The focus on firms with complete data only may introduce a sample bias, because firms with specific characteristics are more likely to enter in our sample. However, in our case, this bias is marginal given the high percentage of retained companies from each country. Moreover, as Andrén and Jankensgård (2015) state, balancing the panel has an important benefit as it allows the possibility to perform different robustness checks.

The investment dynamics (*inv*) is calculated as the growth rate of fixed assets. The liquidity ratios (general liquidity ratio – *lr* and current ratio – *cr*), as well as the profitability ratios (Return on Equity – *roe* and Return on Assets – *roa*) are extracted from AMADEUS database, while the capitalization ratio (*cap*) is equivalent with the capital to total assets ratio.

Table 3 presents the results of panel unit root tests for all variables and countries. With a small exception (the $t^*$ test indicates the absence of stationarity for investment and capitalization in the case of Italy), all variables are stationary and GMM models may be tested.

Table 3. Panel unit root tests

|  | Levin–Lin–Chu | Im–Pesaran–Shin | ADF–Fisher | PP–Fisher |
|---|---|---|---|---|
|  | *t** | *W-stat* | *Chi-square* | *Chi-square* |
| France |  |  |  |  |
| *inv* | -178.48*** | -26.687*** | 1283.5*** | 1832.4*** |
| *cap* | -59.872*** | -4.8567*** | 826.21*** | 1139.2*** |
| *lr* | -29.625*** | -4.2284*** | 938.34*** | 1266.6*** |
| *cr* | -136.49*** | -8.8148*** | 875.02*** | 1255.3*** |
| *roe* | -95.209*** | -13.785*** | 1162.3*** | 1672.9*** |
| *roa* | -93.703*** | -14.462*** | 1112.0*** | 1577.2*** |
| Italy |  |  |  |  |
| *inv* | -10523*** | -3696.9*** | 1830.4*** | 1708.2*** |
| *cap* | -633.61*** | -40.561*** | 860.73*** | 1455.6*** |
| *lr* | -34.530*** | -3.5804*** | 871.94*** | 1191.2*** |
| *cr* | -25.908*** | -2.1644** | 872.10*** | 1042.6*** |
| *roe* | -55.071*** | -11.468*** | 1051.9*** | 1635.9*** |
| *roa* | -43.487*** | -8.1827*** | 971.91*** | 1396.2*** |
| Spain | | | | |
| inv | 504.00 | -33.357*** | 1882.1*** | 2807.2*** |
| cap | 0.2664 | -11.625*** | 1053.3*** | 1270.6*** |
| lr | -38.522*** | -3.9996*** | 1179.6*** | 1581.3*** |
| cr | -33.441*** | -3.9028*** | 1226.2*** | 1498.1*** |
| roe | -254.89*** | -19.882*** | 1409.3*** | 2367.1*** |
| roa | -214.84*** | -14.507*** | 1327.7*** | 2044.8*** |

*Notes: (i) \*, \*\*, \*\*\*, mean stationarity significant at 10 %, 5 % and 1 %; (ii) For all the tests, the null hypothesis is that the panel contains a unit root; (iii) Probabilities for Fisher tests are computed using an asymptotic Chi-square distribution, while the other tests assume asymptotic normality.*

*3.2. Methodology*

Classical panel data analyses investigating the role of firms' financial performance on their investment behavior usually use fixed effects models to avoid the omitted variables bias. Therefore, along with previous studies, we draw first on a panel fixed effects model (Eq. 1).

$$Y_{i,t} = \alpha_0 + \alpha_1 X_{i,t} + \beta_i + \varepsilon_{i,t} \tag{1}$$

where: $Y_{it}$ is the dependent variable (*inv*); $\alpha_0$ is the intercept; $\beta_i$ represents all the stable characteristics of firms from each country; $X_{it}$ represents the vector of independent financial performance variables; $\alpha_1$ are the coefficients; $\varepsilon_{i,t}$ is the error term.

Given the fact that our sample has a N>T structure (the number of companies is much higher than the number of periods), we also test a random model (Eq. 2), which controls for all stable covariates (Allison and Waterman, 2002). To select between these two static models, a Hausman test is performed.

$$Y_{i,t} = \alpha_0 + \alpha_1 X_{i,t} + \beta_i + \mu_{i,t} + \varepsilon_{i,t} \tag{2}$$

where: μ represents between-entity errors; $\varepsilon_{i,t}$ are the within-entity errors.

The results of the classic static models might be affected by an endogeneity bias. While the firms' financial performance influences the investment behavior in the wine industry, we can also expect that an increase in investment will have a negative impact on liquidity and profitability in the short-run, and an opposite effect in the long-run. Further, static models do not account for dynamics, where changes in explicative variables influence the dependent variables after a time adjustment, that is, in the long-run. Therefore, we address the endogeneity issue applying a GMM approach. We first resort to the dynamic-GMM estimator of Arellano and Bond (1991):

$$\Delta investment_{i,t} = \sum_{j=t-p}^{t-1} \vartheta_j \Delta investment + \alpha_1 \Delta capitalisation_{i,t} + \alpha_2 \Delta liquidity_{i,t} +$$
$$\alpha_3 \Delta profitability_{i,t} + \Delta \mu_{i,t} + \Delta u_{i,t} \tag{3}$$

where: $\vartheta$ is the first lag of investment dynamics; $\mu_{i,t}$ and $u_{i,t}$ are the error terms which vary over both firms and time; $\alpha$ are the coefficients of the explanatory variables.

However, for large N and small T samples, the system-GMM might have better properties (see Blundell and Bond, 1998), because in the case of difference-GMM estimator, lagged levels of regressors are considered poor instruments and $\Delta investment_{i,t}$ might be still correlated with $\Delta u_{i,t}$. The system-GMM estimator implies a system of two simultaneous equations, one in level and one in first difference. In this case, both lagged first differences and lagged levels of variables act as instruments.

Both GMM estimators might suffer from the proliferation of instruments and a Sargan test is used for over-identifying restrictions related to instruments. However, the Sargan test is not powerful enough in the presence of too many instruments. Therefore, a Hansen test statistic should be used if nonsphericity is suspected in the errors, which requires robust error correction (Roodman, 2009).

In conclusion, the two GMM estimators we use (difference- and system-GMM) serve as different tools for testing the robustness of our findings. In addition, we also check the robustness by using a two-steps estimator instead of the default one-step. The two-steps estimator requires robust errors and in this case, the standard covariance matrix is robust to panel-specific autocorrelation and heteroscedasticity. Further, in the two-steps approach the number of parameters does not grow with the number of estimated regressors in the nonlinear GMM step. The autocorrelation issue is checked with the Arellano–Bond tests (AR(1) and AR(2)) for autocorrelation, applied to differenced residuals. While the AR(1) process usually rejects the null hypothesis of no autocorrelation, the AR(2) test is more important as it helps detecting the autocorrelation in levels.

## 4. Empirical findings

This section presents the results obtained for each country retained into analysis. The findings of static estimators are presented in Appendix A and serve as reference for potential comparisons with similar researches. According to the fixed and random effects models, there is no significant influence of firms' financial performance on their investment behavior in the case of France and Italy. However, the capitalization and liquidity negatively affect the investment dynamics in Spain, while the profitability level has an opposite effect.

In what follows, we focus on the dynamic estimators' results, and we present the empirical findings for each country. For each estimator, four different models are tested (Models 1-4), resulting from an alternative use of liquidity ratios (*lr* and *cr*) and profitability ratios (*roe* and *roa*). While liquidity and profitability are considered endogenous variables, the capitalization ratio is included in estimations strictly as exogenous variable. There is no theoretical intuition that shows a direct increase or decrease in the level of capitalization, following an increase in the level of investment.

*4.1. Results for France*

In the case of France, the first set of estimations (one-step results) shows in general robust findings between difference- and system-GMM estimators (Table 4). As expected, in all the cases the capitalization level negatively influences the investment dynamics. This result states that an increase of the capitalization ratio might be made in the detriment of an increase in investments. While the liquidity is not important for the investment dynamics, the profitability has a positive influence, as expected. However, this last result is influenced by the way the profitability is measured, a significant influence being reported only in the case of *roe*.

Table 4. GMM results for France (one-step results, GMM errors)

|  | difference-GMM | | | | system-GMM | | | |
|---|---|---|---|---|---|---|---|---|
|  | Model 1 | Model 2 | Model 3 | Model 4 | Model 1 | Model 2 | Model 3 | Model 4 |
| c | 20.31*** | 20.49*** | 21.49*** | 21.73*** | 14.26*** | 14.24*** | 16.16*** | 16.05*** |
| lag(1) | 0.000* | 0.000 | 0.000* | 0.000 | 0.001* | 0.000 | 0.001* | 0.000 |
| cap | -2.462*** | -2.268*** | -2.447*** | -2.254*** | -1.846*** | -1.692*** | -1.884*** | -1.731*** |
| lr | -0.533 | -0.364 |  |  | -0.138 | -0.284 |  |  |
| cr |  |  | -0.725 | -0.659 |  |  | -0.687 | -0.730 |
| roe | 0.666*** |  | 0.671*** |  | 1.174*** |  | 1.196*** |  |
| roa |  | 0.539 |  | 0.556 |  | 2.205*** |  | 2.261*** |
| observations |  | 1,986 | | | | 2,317 | | |
| groups |  | 331 | | | | 331 | | |
| instruments |  | 94 | | | | 59 | | |
| Sargan over- | 721.4 | 724.3 | 719.8 | 722.5 | 885.4 | 896.9 | 886.4 | 898.5 |
| identification | [0.00] | [0.00] | [0.00] | [0.00] | [0.00] | [0.00] | [0.00] | [0.00] |

*Notes: (i) lag(1) is the first lag of the dependent variable; (ii) capitalization is considered strictly exogenous while liquidity and profitability are endogenous variables; (iii) \*, \*\*, \*\*\* means significance at 10 %, 5 % and 1 %; (iv) inv – investment dynamics, cap – capitalization ratio, lr – liquidity ratio, cr – current ratio, roe – return on equity, roa – return on assets.*

The Sargan test shows, nevertheless, that these findings might be affected by the proliferation of instruments. Therefore, in the second part we have performed a two-steps estimation, where the number of maximum lags for the dependent variable is set at one and for the explanatory variable at two. In this case, the results do not indicate a significant influence of financial performances on the investment dynamics (Table 5). The findings are similar for both estimators and for all the models, and in agreement with the static analysis (Appendix A). Moreover, in this case, the Arellano–Bond tests show no autocorrelation problem, while the Sargan and Hansen tests indicate that the instruments are well identified.

We thus conclude that in the case of France, the capitalization negatively impacts the investment dynamics, while the profitability has a positive impact. The liquidity has no significant influence on investment. However, these findings might be influenced by the over-identification of instruments and are not confirmed by the two-steps estimation, which puts into question their robustness.

Table 5. GMM results for France (two-steps results, robust errors)

|  | difference-GMM | | | | system-GMM | | | |
|---|---|---|---|---|---|---|---|---|
|  | Model 1 | Model 2 | Model 3 | Model 4 | Model 1 | Model 2 | Model 3 | Model 4 |
| c | 12.48 | 11.55 | 13.52* | 12.56 | 3.621 | 1.520 | 7.926 | 6.261 |
| lag(1) | 0.000*** | 0.000*** | 0.000*** | 0.000*** | 0.073 | 0.073 | 0.031 | 0.037 |
| cap | -2.123 | -1.747 | -2.113 | -1.729 | -0.015 | -0.002 | -0.026 | -0.021 |
| lr | -0.397 | -0.203 |  |  | -2.430 | -1.722 |  |  |
| cr |  |  | -0.550 | -0.453 |  |  | -1.593 | -1.564 |
| roe | 0.654 |  | 0.668 |  | 0.237 |  | -0.007 |  |
| roa |  | 0.536 |  | 0.603 |  | 0.880 |  | 0.456 |
| observations |  | 1,986 | | | | 2,317 | | |
| groups |  | 331 | | | | 331 | | |
| instruments |  | 94 | | | | 32 | | |
| Arellano-Bond test AR(1) | -1.339 [0.18] | -1.325 [0.18] | -1.340 [0.18] | -1.326 [0.18] | -1.320 [0.18] | -1.360 [0.17] | -1.330 [0.18] | -1.350 [0.17] |
| Arellano-Bond test AR(2) | -0.447 [0.65] | -0.143 [0.88] | -0.474 [0.63] | -0.169 [0.86] | 0.310 [0.75] | 0.460 [0.64] | -0.020 [0.98] | 0.080 [0.93] |
| Sargan over-identification |  |  |  |  | 7.170 [1.00] | 10.29 [0.99] | 19.70 [0.84] | 18.40 [0.89] |
| Hansen over-identification |  |  |  |  | 27.62 [0.43] | 24.52 [0.60] | 22.66 [0.70] | 21.74 [0.75] |

*Notes: Similar to Table 4.*

## 4.2. Results for Italy

In the case of the Italian wine industry, the default one-step estimation shows no significant influence of financial performance on investment dynamics, except for the liquidity ratios for the system-GMM approach. Table 6 shows no significant impact of capitalization and profitability, while the Sargan over-identification test indicates a proliferation of instruments issue.

Table 6. GMM results for Italy (one-step result, GMM errors)

|  | difference-GMM | | | | system-GMM | | | |
|---|---|---|---|---|---|---|---|---|
|  | Model 1 | Model 2 | Model 3 | Model 4 | Model 1 | Model 2 | Model 3 | Model 4 |
| c | 24.24*** | 27.19*** | 16.26** | 23.48*** | 7.021 | 9.288 | -1.830 | 1.887 |
| lag(1) | -0.000 | -0.000 | -0.000 | -0.000 | -0.000 | -0.001 | -0.000 | -0.001 |
| cap | -0.127 | -0.177 | 0.075 | -0.034 | -0.076 | -0.126 | 0.087 | 0.005 |
| lr | -3.858 | -5.958 |  |  | 12.71*** | 10.25*** |  |  |
| cr |  |  | 1.709 | -1.953 | -0.044 |  | 12.23*** | 9.779*** |
| roe | -0.006 |  | 0.013 |  |  |  | -0.031 |  |
| roa |  | -0.617 |  | -0.399 |  | 0.655 |  | 1.149 |
| observations |  | 2,010 | | | | 2,345 | | |
| groups |  | 335 | | | | 335 | | |
| instruments |  | 94 | | | | 59 | | |
| Sargan over-identification | 615.7 [0.00] | 635.3 [0.00] | 489.0 [0.00] | 546.9 [0.00] | 741.1 [0.00] | 777.7 [0.00] | 601.7 [0.00] | 671.5 [0.00] |

*Notes: Similar to Table 4.*

These findings are this time confirmed by the two-steps estimations with robust errors and we notice once again the lack of a significant influence of firms' financial performance on their investment dynamics in Italy (Table 7). As in the case of France, the two-steps estimations for Italy do not present autocorrelation or over-identification problems.

Table 7. GMM results for Italy (two-step results, robust errors)

|  | difference-GMM | | | | system-GMM | | | |
| --- | --- | --- | --- | --- | --- | --- | --- | --- |
|  | Model 1 | Model 2 | Model 3 | Model 4 | Model 1 | Model 2 | Model 3 | Model 4 |
| c | 13.29*** | 15.69*** | 6.902 | 13.04*** | 12.98*** | 13.70*** | 12.89*** | 13.60 |
| lag(1) | -0.000 | -0.000 | -0.000 | -0.000 | 0.016 | 0.003 | 0.033* | 0.007 |
| cap | -0.099 | -0.111 | 0.083 | -0.027 | -0.086 | -0.105 | -0.089 | -0.081 |
| lr | -3.952 | -6.044** |  |  | -1.548 | -1.127 |  |  |
| cr |  |  | 1.675 | -1.969 |  |  | -0.956 | -0.738 |
| roe | -0.006 |  | 0.008 |  | 0.026 |  | 0.057 |  |
| roa |  | -0.745 |  | -0.491 |  | -0.199 |  | -0.115 |
| observations |  | 2,010 | | |  | 2,345 | | |
| groups |  | 335 | | |  | 335 | | |
| instruments |  | 94 | | |  | 59 | | |
| Arellano-Bond test AR(1) | -1.716 [0.08] | -1.715 [0.08] | -1.717 [0.08] | -1.716 [0.08] | -1.750 [0.08] | -1.720 [0.08] | -1.750 [0.08] | -1.720 [0.08] |
| Arellano-Bond test AR(2) | 0.321 [0.74] | 0.161 [0.87] | 0.686 [0.49] | 0.454 [0.64] | 0.850 [0.39] | 0.610 [0.54] | 1.150 [0.25] | 0.730 [0.46] |
| Sargan over-identification |  |  |  |  | 3.260 [1.00] | 4.280 [1.00] | 2.580 [1.00] | 3.690 [1.00] |
| Hansen over-identification |  |  |  |  | 30.99 [0.27] | 27.55 [0.43] | 31.32 [0.25] | 29.78 [0.32] |

*Notes: Similar to Table 4.*

## 4.3. Results for Spain

The first set of results recorded for Spain (Table 8) shows that, in the case of a one-step classical estimation, the capitalization ratio has a significant and negative impact on investment for all tested models, while the profitability has a positive impact, regardless the way profitability is computed. For firms acting in Spain, we notice that liquidity negatively influences the investment behavior. That is, firms that decide to increase their liquidity accept a reduction in the investment growth rate and conversely, the increase of investment is made in the detriment of the liquidity level. This result can be explained by the fact that Spanish wine companies might use their own funds with predilection, to finance the investment opportunities.

The two-steps estimation partially confirms the one-step findings, although the significance of results decreases (Table 9). For the difference-GMM estimator, for all the models, we notice a negative impact of capitalization and liquidity, and a positive influence of profitability on the investment dynamics. However, for the system-GMM estimator, the significance of liquidity and profitability's coefficients is no longer recorded.

Table 8. GMM results for Spain (one-step results, GMM errors)

| | difference-GMM | | | | system-GMM | | | |
|---|---|---|---|---|---|---|---|---|
| | Model 1 | Model 2 | Model 3 | Model 4 | Model 1 | Model 2 | Model 3 | Model 4 |
| c | 14.02*** | 12.51*** | 14.79*** | 13.26*** | 16.10*** | 15.35*** | 16.69*** | 15.91*** |
| lag(1) | 0.052** | 0.054*** | 0.050*** | 0.052*** | 0.023 | 0.020 | 0.023 | 0.020 |
| cap | -0.236** | -0.193* | -0.217** | -0.174* | -0.336*** | -0.320*** | -0.319*** | -0.303*** |
| lr | -1.580*** | -1.565*** | | | -0.940*** | -0.912*** | | |
| cr | | | -1.137*** | -1.128*** | | | -0.770*** | -0.752*** |
| roe | 0.067* | | 0.067* | | 0.075** | | 0.075** | |
| roa | | 0.325 | | 0.326 | | 0.400* | | 0.411* |
| observations | | | 2,652 | | | | 3,094 | |
| groups | | | 442 | | | | 442 | |
| instruments | | | 94 | | | | 59 | |
| Sargan over-identification | 215.7 [0.00] | 202.0 [0.00] | 211.2 [0.00] | 199.8 [0.00] | 190.0 [0.00] | 228.2 [0.00] | 185.8 [0.00] | 222.9 [0.00] |

*Notes: Similar to Table 4.*

If in the case of the one-step estimators the Sargan test indicates an instrument over-identification problem, in the case of the two-steps estimators (Table 9), the Sargan and Hansen tests show that instruments are well identified, and the autocorrelation test shows no autocorrelation bias, especially for the system-GMM specification.

Table 9. GMM results for Spain (two-steps results, robust errors)

| | difference-GMM | | | | system-GMM | | | |
|---|---|---|---|---|---|---|---|---|
| | Model 1 | Model 2 | Model 3 | Model 4 | Model 1 | Model 2 | Model 3 | Model 4 |
| c | 6.610*** | 7.872*** | 8.229*** | 8.831*** | 5.624*** | 4.469** | 5.614*** | 4.182** |
| lag(1) | 0.065*** | 0.065*** | 0.062*** | 0.062*** | 0.019 | -0.095 | 0.009 | -0.068 |
| cap | -0.103 | -0.149* | -0.119* | -0.140* | -0.079*** | -0.038 | -0.077*** | -0.030 |
| lr | -1.499* | -1.528* | | | -0.174 | -0.167 | | |
| cr | | | -1.036** | -1.062** | | | -0.053 | -0.105 |
| roe | 0.062** | | 0.057 | | -0.009 | | 0.023 | |
| roa | | 0.437* | | 0.361 | | 0.699 | | 0.820 |
| observations | | | 2,652 | | | | 3,094 | |
| groups | | | 442 | | | | 442 | |
| instruments | | | 94 | | | | 32 | |
| Arellano-Bond test AR(1) | -3.171 [0.00] | -3.179 [0.00] | -3.153 [0.00] | -3.165 [0.00] | -2.080 [0.03] | -2.100 [0.03] | -2.250 [0.02] | -2.440 [0.01] |
| Arellano-Bond test AR(2) | 1.687 [0.09] | 1.628 [0.10] | 1.550 [0.12] | 1.515 [0.12] | 0.210 [0.83] | -0.059 [0.55] | 0.170 [0.86] | -0.470 [0.64] |
| Sargan over-identification | | | | | 55.01 [0.00] | 59.66 [0.00] | 46.77 [0.02] | 52.48 [0.00] |
| Hansen over-identification | | | | | 19.92 [0.83] | 26.70 [0.48] | 21.57 [0.75] | 28.62 [0.38] |

*Notes: Similar to Table 4.*

## 5. Summary of results, comparisons and policy implications

This section presents a short overview of the empirical findings in a comparative manner and discusses different financial management strategies that seems to be implemented by the firms acting in the wine industry from the largest worldwide producers. Table 10 shows that our empirical findings are in general robust to different estimators and models we have used but are sensitive to the way we address the proliferation of instrument issue.

Table 10. Results' centralization

| *investment dynamics* | difference-GMM | | system-GMM | |
|---|---|---|---|---|
| | one-step | two-steps | one-step | two-steps |
| France | | | | |
| *capitalization* | N | - | N | - |
| *liquidity* | - | - | - | - |
| *profitability* | P | - | P | - |
| Italy | | | | |
| *capitalization* | - | - | - | - |
| *liquidity* | - | - | P | - |
| *profitability* | - | - | - | - |
| Spain | | | | |
| *capitalization* | N | N | N | N |
| *liquidity* | N | N | N | - |
| *profitability* | P | P | P | - |
| *Notes: (i) 'P / N' means positive / negative significant influence; (ii) '-' indicates no significant influence.* | | | | |

We can notice that, in the case of Italy, the financial performance of wine industry companies does not influence their investment behavior. That is, the investment decision is based on other factors (e.g. market conditions), and we may suppose these companies extend their production capacity by accessing external funds, in the detriment of internal sources. This result might also indicate a lack of inertia regarding the investment dynamics in the aftermath of the recent global financial crisis.

For the French wine companies, the degree of capitalization and the level of profitability represent reliable factors which influence their investment dynamics. In general, the profitability favors the investment decision, while a trade-off is recorded between investment and capitalization. It appears that internal funds play their role in the investment behavior, although the results in case of France are not very robust.

In the case of Spanish wine companies, we notice an important role of financial performance in influencing their investment behavior. On the one hand, the capitalization and liquidity ratios have a negative influence on the investment dynamics. On the other hand, a higher profitability represents a prerequisite for increasing the investment level. These findings are quite robust and show that Spanish managers from the wine industry prefer the internal funds to extend their business. The results reported for Spain indicate the existence of a trade-off between capitalization and liquidity on the one hand, and investment dynamics on the other hand. Moreover, these results confirm the potential trade-off between liquidity and profitability underlined by previous researches.

## 6. Conclusions

The purpose of this paper was to investigate how firms' investment behavior is influenced by their financial performance. With a focus on the wine industry from the largest EU producers namely France, Italy and Spain, we use firm-level data for a large set of companies to perform this investigation. Our panel data analysis covers the post-crisis period (2007 to 2014) and relies on dynamic model specifications.

The findings show different investment strategies for firms located in these countries. It appears that the investment behavior of Italian firms is not influenced by their financial performance. In addition, in the case of French companies, only the capitalization and the profitability ratio are important for the investment decision, while the influence of liquidity is insignificant. However, these results are partially robust and might be affected by the over-identification of the instruments used in the analysis. Finally, interesting and robust results are reported for Spanish firms. We show that the financial performance of wine companies is very important for their investment behavior. If a negative impact is recorded in the case of capitalization and liquidity, a positive influence is noticed for the profitability level. This means that the profits are usually re-invested by Spanish companies, and that internal funds are preferred by managers to sustain their investment decision. These findings support the growing importance of the Spanish wine industry at global level and have noteworthy policy implications for financial managers acting in these companies, as well as for the national authorities interested in the development and increased performance of the wine sector.

**Acknowledgements**

This work was supported by a grant of the Romanian National Authority for Scientific Research and Innovation, CNCS – UEFISCDI, project number PN-III-P1-1.1-TE-2016-0142.

## References

Abel, A. B. (1983). Optimal investment under uncertainty. American Economic Review, 73, 228–233.

Acharya, V. V., Almeida, H., and Campello, M. (2007). Is cash negative debt? A hedging perspective on corporate financial policies. Journal of Financial Intermediation, 16, 515–554.

Ahn, S., Denis , D., and Denis, D. (2006). Leverage and investment in diversified firms. Journal of Financial Economics, 79, 317–337.

Aidogan, A. (2003). How sensitive is investment to cash flow when financing is frictionless? Journal of Finance, 58, 707–722.

Aivazian, V. A, Ge, J., and Qiu, J. (2005). The impact of leverage on firm investment: Canadian evidence. Journal of Corporate Finance, 11, 277–291.

Ajide, F.M. (2017). Firm-specific, and institutional determinants of corporate investments in Nigeria. Future Business Journal, 3,107–118.

Albulescu, C. T., Miclea, Ș., Suciu, S. S., and Tămăşilă, M. (2017). Firm-level investment in the extractive industry from CEE countries: the role of macroeconomic uncertainty and internal conditions. Eurasian Business Review, doi:10.1007/s40821-017-0079-3.

Alex, A. C., Hong, C., Dayong, Z., and David, G. D. (2013).The impact of shareholding structure on firm investment: Evidence from Chinese listed companies. Pacific-Basin Finance Journal, 25, 85–100.

Allison, P. D., and Waterma, R. P. (2002). Fixed–effects negative binomial regression models. Sociological Methodology, 32, 247–265.

Almeida, H., Campello, M. and Weisbach, M. S. (2011). Corporate financial and investment policies when future financing is not frictionless. Journal of Corporate Finance, 17, 675–693.

Ameer, R. (2014). Financial constraints and corporate investment in Asian Countries. Journal of Asian Economics, 33, 44–55.

Anderson, K., Norman, D., and Wittwer, G. (2003). Globalisation of the world's wine markets. World Economics, 26, 659–687.

Anderson, K. and Wittwer, G. (2017). U.K. and global wine markets by 2025, and implications of Brexit. Journal of Wine Economics, 12, 221–251.

Andrén, N., and Jankensgård, H. (2015). Wall of cash: The investment-cash flow sensitivity when capital becomes abundant. Journal of Banking and Finance, 50, 204–213.

Arellano, M., and Bond, S. R. (1991). Some tests of specification for panel data: Monte Carlo evidence and an application to employment equations. Review of Economic Studies, 58, 277–297.

Arslan-Ayaydin, Ö., Florackis, C., and Ozkan, A. (2014). Financial flexibility, corporate investment and performance: evidence from financial crises. Review of Quantitative Finance and Accounting, 42, 211–250.

Baum, C. F., Caglayan, M., and Talavera, O. (2008). Uncertainty determinants of firm investment. Economics Letters, 98, 282–287.

Baum, C. F., Caglayan, M., and Talavera, O. (2010). On the investment sensitivity of debt under uncertainty. Economics Letters, 106, 25–27.

Bernanke, B. S. (1983). Irreversibility, uncertainty, and cyclical investment. Quarterly Journal of Economics, 98, 85–106.

Black, E., Legoria, J., and Sellers, K. (2000). Capital investment effects of dividend imputation. Journal of the American Taxation Association, 22, 40–59.

Blundell, R. W., and Bond, S. R. (1998). Initial conditions and moment restrictions in dynamic panel data models. Journal of Econometrics, 87, 115–143.

Bokpin, G. A., and Onumah, J. M. (2009). An empirical analysis of the determinants of corporate investment decisions: Evidence from emerging market firms. International Research Journal of Finance and Economics, 33, 134–141.

Calcagnini, G., and Iacobucci, D. (1997). Small firm investment and financing decisions: an option value approach. Small Business Economics, 9, 491–502.

Chen, S-S., Chung, T-Y., and Chung, L-I. (2001). Investment opportunities, free cash flow and stock valuation effects of corporate investments: The case of Taiwanese investments in China. Review of Quantitative Finance and Accounting, 16, 299–310.

Chen, T., Xie, L., and Zhang, Y. (2017). How does analysts' forecast quality relate to corporate investment efficiency? Journal of Corporate Finance, 43, 217–240.

Colombo, M.G., Croce, A. and Guerini, M. (2013). The effect of public subsidies on firms' investment–cash flow sensitivity: Transient or persistent? Research Policy, 42, 1605–1623.

Danielson, M. G., and Scott, J. A. (2007). A note on agency conflicts and the small firm investment decision. Journal of Small Business Management, 45, 157–175.

Datta, D, and Agarwal, B. (2014). Corporate investment behaviour in India during 1998–2012: bear, bull and liquidity phase. Paradigm, 18, 87–102.

Farla, K. (2014). Determinants of firms' investment behaviour: a multilevel approach. Applied Economics, 46, 4231–4241.

Fazzari, S., Hubbard, G., and Petersen, B. (1988). Financing constraints and corporate investment. Brookings Papers on Economics Activity, 1, 141–195.

Gamba, A., and Triantis, A. (2008). The value of financial flexibility. Journal of Finance, 63, 2263–2296.

Gatchev, V. A., Spindt, P. A., and Tarhan,V. (2009). How do firms finance their investments?: The relative importance of equity issuance and debt contracting costs. Journal of Corporate Finance, 15, 179–195.

Gertler, M., and Gilchrist, S. (1994). Monetary policy, business cycles, and the behaviour of small manufacturing firms. Quarterly Journal of Economics, 109, 309–340.

Gilchrist, S., and Himmelberg, C. (1995). Evidence on the role of cash flow for investment. Journal of Monetary Economics, 36, 541–572.

Glover, B., and Levine, O. (2015). Uncertainty, investment, and managerial incentives. Journal of Monetary Economics, 69, 121–137.

Hall, R., and Jorgenson, D. (1967). Tax policy and investment behaviour. American Economic Review, 57, 391–414.

Hartman, R. (1972). The effects of price and cost uncertainty on investment. Journal of Economic Theory, 5, 258–266.

Hirth, S., and Viswanatha,M. (2011). Financing constraints, cash-flow risk, and corporate investment. Journal of Corporate Finance, 17, 1496–1509.

Jensen, M., and Meckling, W. (1976). Theory of the firm: Managerial behavior, agency costs and ownership structure. Journal of Financial Economics, 3, 305–360.

Jeon, H., and Nishihara, M. (2014). Macroeconomic conditions and a firm's investment decisions. Finance Research Letters, 11, 398–409.

Jugurnath, B., Stewart, M, and Brooks, R. (2008). Dividend taxation and corporate investment: a comparative study between the classical system and imputation system of dividend taxation in the United States and Australia. Review of Quantitative Finance and Accounting, 31, 209–224.

Kang, T., Baek, C. and Lee, J-d. (2017). The persistency and volatility of the firm R&D investment: Revisited from the perspective of technological capability. Research Policy, 46, 1570–1579.

Kim, T. N. (2014). The impact of cash holdings and external financing on investment-cash flow sensitivity. Review of Accounting and Finance, 13, 251–273.

Koo, J., and Maeng, K. (2006). Foreign ownership and investment: evidence from Korea. Applied Economics, 38, 2405–2414.

Lang, L., Ofek, E., and Stulz, R. M. (1996). Leverage, investment, and firm growth. Journal of Financial Economics, 40, 3–29.

Leary, M. T., and Roberts, M. R. (2014). Do peer firms affect corporate financial policy? Journal of Finance, 69, 139–178.

Lyandres, E. (2006). Capital structure and interaction among firms in output markets: Theory and evidence. Journal of Business, 79, 2381–2421.

Maçãs Nunes, P., Mendes, S., and Serrasqueiro, Z. (2012). SMEs' investment determinants: empirical evidence using quantile approach. Journal of Business Economics and Management, 13, 866–894.

Mavruk, T., and Carlsson, E. (2015). How long is a long-term-firm investment in the presence of governance mechanisms? Eurasian Business Review, 5, 117–149.

Modiglian, F., and Miller, M. (1958). The cost of capital, corporation finance and the theory of investment. American Economic Review, 48, 261–297.

Morck, R. (2003). Why some double taxation might make sense: The special case of intercorporate dividends. Working Paper Series National Bureau of Economic Research 9651.

Morrison, A., and Rabellotti, R. (2017). Gradual catch up and enduring leadership in the global wine industry. Research Policy, 46, 417–430.

Mulier, K., Schoors, K., and Merlevede, B. (2016). Investment-cash flow sensitivity and financial constraints: Evidence from unquoted European SMEs. Journal of Banking and Finance, 73, 182–197.

Outreville, J. F., and Hanni, M. (2013). Multinational firms in the world wine industry: an investigation into the determinants of most-favoured locations. Journal of Wine Research, 24, 128–137.

Park, K., Yang, I., and Yang, T. (2017). The peer-firm effect on firm's investment decisions. North American Journal of Economics and Finance, 40, 178–199.

Pérez-Orive, A. (2016). Credit constraints, firms' precautionary investment, and the business cycle. Journal of Monetary Economics, 78, 112–131.

Perić, M., and Đurkin, J. (2015). Determinants of investment decisions in a crisis: perspective of Croatian small firms. Management, 20, 115–133.

Pindyck, R. S. (1988). Irreversible investment, capacity choice, and the value of the firm. American Economic Review, 78, 969–985.

Rizzo, A. M. (2019). Competition policy in the wine industry in Europe. Journal of Wine Economics, 14, 90–113.

Roodman, D. (2009). How to do xtabond2: an introduction to difference and system GMM in Stata. The Stata Journal, 9, 86–136.

Stickney, C., and McGee, V. (1982). Effective corporate tax rates: the effect of size, capital intensity, leverage, and other factors. Journal of Accounting and Public Policy, 1, 125–152

Suto, M. (2003). Capital structure and investment behaviour of Malaysian firms in the 1990s: a study of corporate governance before the crisis. Corporate Governance: An International Review, 11, 25–39.

Vermoesen, V., Deloof, M. and Laveren, E. (2013). Long-term debt maturity and financing constraints of SMEs during the Global Financial Crisis. Small Business Economics, 41, 433–448.

Vithessonthi, C., Schwaninger, M., and Müller, M.O. (2017). Monetary policy, bank lending and corporate investment. International Review of Financial Analysis, 50, 129–142.

Yu, X., Dosi, G., Grazzi, M. and Lei, J. (2017). nside the virtuous circle between productivity, profitability, investment and corporate growth: An anatomy of Chinese industrialization. Research Policy, 46, 1020–1038.

# Appendix

*Appendix A. Static panel data analysis*

Table A1. Results of fixed and random effect estimators for France

| France | | Model 1 | | Model 2 | | Model 3 | | Model 4 | |
|---|---|---|---|---|---|---|---|---|---|
| Variables | Models | Fixed effects | Random effects | Fixed effects | Random effects | Fixed effects | Random effects | Fixed effects | Random effects |
| $c$ | | 109.0 | 107.4 | 92.32 | 82.41 | 107.6 | 118.8 | 90.37 | 93.06 |
| | | (106) | (68.73) | (108) | (69.39) | (108) | (71.34) | (111) | (72.18) |
| $cap$ | | -2.158 | -0.579 | -4.123 | -0.338 | -2.173 | -0.535 | -4.155 | -0.298 |
| | | (13.49) | (4.287) | (13.45) | (4.290) | (13.49) | (4.277) | (13.44) | (4.281) |
| $lr$ | | -1.639 | -10.06 | -3.861 | -9.031 | | | | |
| | | (37.71) | (26.49) | (37.77) | (26.52) | | | | |
| $cr$ | | | | | | -0.315 | -9.197 | -1.157 | -8.406 |
| | | | | | | (20.69) | (15.33) | (20.71) | (15.34) |
| $roe$ | | -7.237 | -6.188* | | | -7.242 | -6.220* | | |
| | | (4.716) | (3.473) | | | (4.714) | (3.473) | | |
| $roa$ | | | | 1.431 | -1.416 | | | 1.374 | -1.520 |
| | | | | (16.96) | (10.89) | | | (16.95) | (72.18) |
| Hausman test (recommended) | | Prob>chi2 = 0.97 (Random) | | Prob>chi2 = 0.98 (Random) | | Prob>chi2 = 0.91 (Random) | | Prob>chi2 = 0.93 (Random) | |
| *Notes*: (i) *, **, *** means significance at 10 %, 5 % et 1 %; (ii) Standard errors are reported in brackets. | | | | | | | | | |

Table A2. Results of fixed and random effect estimators for Italy

| Italy | | Model 1 | | Model 2 | | Model 3 | | Model 4 | |
|---|---|---|---|---|---|---|---|---|---|
| Variables | Models | Fixed effects | Random effects | Fixed effects | Random effects | Fixed effects | Random effects | Fixed effects | Random effects |
| $c$ | | 71.71 | 45.86 | 73.77 | 47.18 | 73.94 | 49.99 | 75.99 | 50.82 |
| | | (45.56) | (35.54) | (46.70) | (35.85) | (52.33) | (38.61) | (53.30) | (38.83) |
| $cap$ | | -0.388 | -0.829 | -0.364 | -0.839 | -0.387 | -0.867 | -0.363 | -0.873 |
| | | (3.452) | (2.458) | (3.454) | (2.458) | (3.453) | (2.463) | (3.455) | (2.463) |
| $lr$ | | -21.44 | 6.393 | -21.40 | 7.107 | | | | |
| | | (27.28) | (12.51) | (27.28) | (12.74) | | | | |
| $cr$ | | | | | | -14.15 | 1.522 | -14.12 | 1.885 |
| | | | | | | (22.06) | (11.21) | (22.06) | (11.34) |
| $roe$ | | 0.000 | 0.036 | | | 0.009 | 0.042 | | |
| | | (1.043) | (0.734) | | | (1.043) | (0.734) | | |
| $roa$ | | | | -2.483 | -2.221 | | | -2.491 | -1.583 |
| | | | | (12.28) | (7.782) | | | (12.28) | (7.730) |
| Hausman test (recommended) | | Prob>chi2 = 0.71 (Random) | | Prob>chi2 = 0.69 (Random) | | Prob>chi2 = 0.86 (Random) | | Prob>chi2 = 0.85 (Random) | |
| *Notes*: (i) *, **, *** means significance at 10 %, 5 % et 1 %; (ii) Standard errors are reported in brackets. | | | | | | | | | |

Table A3. Results of fixed and random effect estimators for Spain

| Spain | Model 1 | | Model 2 | | Model 3 | | Model 4 | |
|---|---|---|---|---|---|---|---|---|
| Variables \ Models | Fixed effects | Random effects | Fixed effects | Random effects | Fixed effects | Random effects | Fixed effects | Random effects |
| $c$ | 14.50*** | 9.399*** | 13.82*** | 8.747*** | 15.17*** | 9.845*** | 14.49*** | 9.176*** |
| | (2.338) | (35.54) | (2.364) | (1.116) | (2.367) | (1.103) | (2.392) | (38.83) |
| $cap$ | -0.241*** | -0.092*** | -0.225*** | -0.075*** | -0.231*** | -0.086*** | -0.214*** | -0.068** |
| | (0.770) | (0.025) | (0.077) | (0.026) | (0.077) | (0.025) | (0.077) | (0.026) |
| $lr$ | -0.629** | -0.095 | -0.631** | -0.123 | | | | |
| | (0.270) | (0.189) | (0.270) | (0.189) | | | | |
| $cr$ | | | | | -0.597*** | -0.234 | -0.599*** | -0.253* |
| | | | | | (0.206) | (0.144) | (0.206) | (0.144) |
| $roe$ | 0.052* | 0.044* | | | 0.052* | 0.045* | | |
| | (0.029) | (0.026) | | | 0.029) | (0.026) | | |
| $roa$ | | | 0.286 | 0.352** | | | 0.291 | 0.355** |
| | | | (0.180) | (0.139) | | | (0.179) | (0.139) |
| Hausman test (recommend) | Prob>chi2 = 0.00 (Fixed) | | Prob>chi2 = 0.01 (Fixed) | | Prob>chi2 = 0.01 (Fixed) | | Prob>chi2 = 0.01 (Fixed) | |

*Notes*: (i) *, **, *** means significance at 10 %, 5 % et 1 %; (ii) Standard errors are reported in brackets.